\documentclass[%
 reprint,
 amsmath,amssymb,
 aps,
 prd,
floatfix,
twocolumn
]{revtex4-2}

\usepackage{graphicx}
\usepackage{dcolumn}
\usepackage{bm}

\usepackage{xcolor}
\usepackage{graphicx}
\usepackage{epsfig}
\usepackage{slashed}
\usepackage{tabu,multirow}
\usepackage[unicode=true,bookmarks=false,breaklinks=false,pdfborder={0 0 1},colorlinks=true]
 {hyperref}
\hypersetup{
 citecolor=blue,linkcolor=blue,urlcolor=blue}
\begin{document}


\title{A holographic bottom-up description of light nuclide spectroscopy  and stability}

\author{Miguel Angel Martin Contreras}
\email{miguelangel.martin@uv.cl}
\affiliation{
 Escuela de Ciencias \\
 Universidad  Vi\~na del Mar,\\
 Vi\~na del Mar, Chile
}

\affiliation{Corporación Universitaria Minuto de Dios--UNIMINUTO\\
Unidad de Ciencias Básicas\\
Cundinamarca, Colombia}

\author{Alfredo Vega}%
 \email{alfredo.vega@uv.cl}
\affiliation{%
 Instituto de F\'isica y Astronom\'ia, \\
 Universidad de Valpara\'iso,\\
 A. Gran Breta\~na 1111, Valpara\'iso, Chile
}

\author{Saulo Diles}
\email{smdiles@ufpa.br}
\affiliation{Campus Salin\'opolis,\\ Universidade Federal do Par\'a,\\
68721-000, Salin\'opolis, Par\'a, Brazil}

\date{\today}

\begin{abstract}
This work explores a holographic proposal to describe light nuclide spectroscopy by considering extensions to the well-known bottom-up AdS/QCD proposals, the hardwall and softwall models. We also propose an alternative description inspired by the Woods-Saxon potential. We find the static dilaton associated with this potential in this Wood-Saxon-like model. We compute the nuclide spectra finding that, despite their pure AdS/QCD origin, hardwall and softwall, as monoparametric models, have good accuracy and precision since the RMS error is near 11$\%$ and 4 $\%$ respectively. In the case of the Wood-Saxon model, the RMS was around 1$\%$. We also discuss configurational entropy as a tool to categorize which model is suitable to describe nuclides in terms of stability. We found that configurational entropy resembles a stability line, independent from nuclear spin,  for symmetric light nuclides when considering softwall and Wood-Saxon-like models. For the hardwall case, configurational entropy, despite increasing with the constituent number, depends on the nuclear spin. Thus, the Woods-Saxon-like model emerges as the best choice to describe light nuclide spectroscopy in the bottom-up scenario.

\end{abstract}

\maketitle


\section{\label{intro}Introduction}

Holography nowadays is one of the most used tools to describe non-perturbative phenomena. It is not restricted to hadronic physics only (glueballs, mesons, and baryons\cite{Gutsche:2011vb,Gutsche:2015xva,Braga:2015jca,Braga:2016wzx,Braga:2015lck,FolcoCapossoli:2015jnm,MartinContreras:2019kah}). There are applications to heavy ion collisions \cite{Mateos:2011bs,Casalderrey-Solana:2016xfq}, condensed matter \cite{Sachdev:2010ch}, neutron stars \cite{deBoer:2009wk,Arguelles:2019mxh}, or fluid mechanics \cite{Policastro:2002se,Kovtun:2004de, Kovtun:2012rj, Diles:2019uft}. After more than two decades from Maldacena's seminal work \cite{Maldacena:1997re}, holography seems to provide a fruitful soil to develop effective models to approach non-perturbative phenomena. 

Among the non-perturbative systems available in nature, one of the most challenging is the nuclear realm. The nuclear force observed in nucleon systems is a low energy strong force residual approximately, thus implying it is non-perturbative. At the holographic level, this hypothesis allows using bottom-up models to describe such phenomenology. This manuscript will focus on describing the light nuclide spectrum using holographic tools. 

There are significant challenges in describing composite nucleus spectroscopy candidates in holography. The holographic spectrum should describe the nuclear mass as a function of the atomic  $Z$  and mass $A$ numbers. In the case of hadrons, the spectra are organized by defined mass poles, the so-called Regge Trajectories, where each excited state defines a new hadron in the family, as in the AdS/QCD bottom-up radial case  \cite{Boschi-Filho:2002xih, Colangelo:2008us, Braga:2015lck}. In the nuclear case, the excited nuclear states come from transitions defining metastable states that do not differ so much from the nucleus ground state mass. Thus, the holographic nuclear mass spectrum should not have a large mass gap compared to the ground state mass.

Another ingredient to take into account is the holographic dictionary. In the hadronic case, the conformal dimension is connected with the operator that creates hadrons at the boundary. When we consider the twist operator, the conformal dimension associated with the bulk dual field is translated into the hadronic constituent number \cite{Brodsky:2006uqa, deTeramond:2010ge}. In the nuclear case, we expect the same behavior. Since atomic and mass numbers characterize the nucleus, the conformal dimension of the dual bulk field should carry this constituent information. Therefore a sensible AdS/QCD model of light nuclei should capture the constituent nucleon dependence on the dual bulk field conformal dimension.

The discussion on nuclei stability analyses is a more subtle problem. Nuclei are composite states of two different components, i.e., protons and neutrons. Each one is composed of three first-generation valence quarks. So, their baryonic numbers are far from vanishing, and it is not consistent with associating nuclear stability with leptonic annihilation. This fact excludes the possibility of making a stability analysis through electromagnetic decay constants, as mesonic holography literature proves \cite{Braga:2015jca,MartinContreras:2019kah,Ballon-Bayona:2021ibm}. Nuclear stability is associated with the arrangement of nucleons. Thus, looking for the associated information with the different configurations is natural. In this context, the concept of configurational entropy (CE) emerges as a natural observable to account for the nuclear stability \cite{Gleiser:2011di,Gleiser:2013mga}. Fortunately, in the last years, the holographic calculation of configurational entropy has been extensively discussed in the context of bottom-up models \cite{Braga:2017fsb, Braga:2018fyc,Bernardini:2018uuy,Ferreira:2019inu, Ferreira:2019nkz}.

The summary of this work is: in section \ref{sec3} we provide a holographic model for light nuclei in the context of bottom-up models. We discuss how computing the mass spectrum for light-nuclides in the context of the so-called hardwall and softwall models, and a new approach, the hybrid Woods-Saxon model. Section \ref{sec4} addresses nuclear stability from the perspective of configurational entropy. Finally, Section \ref{conc} presents conclusions for this work.

\section{Holographic model for light nuclide}\label{sec3}
In the AdS/QCD \emph{a la} bottom-up, it has been established that colored medium properties observed at the boundary system are encoded into bulk fields and the bulk metric. In nuclear systems, considered as a system of nucleons interacting through a strong residual force, we can extend the hypothesis used in hadronic holography to the nuclear medium.  The intensity of this force, responsible for keeping the nucleus cohered, depends on the mass number $A$ and the atomic number $Z$, i.e., the constituent number. Thus, it is quite natural to extend the bottom-up confinement procedure to mimic the nuclear force.

Following the ideas above, a given nucleus endowed with an atomic number $Z$ and $A$ nucleons ($Z$ protons and $A-Z$ neutrons), i.e., a nuclide, will be dual to normalizable bulk mode. Thus, nuclides in this formulation are considered fundamental objects composed by $A$ nucleons, i.e., we neglect the nuclide inner configuration.  For simplicity, we will focus on symmetric nuclides, i.e., those with $A=2\,Z$. 

Our starting point is to consider the nuclide, defined by a mass number $A$, and the spin $p$, characterized by a $p$-form bulk field $A_p(\zeta,x^\mu)$. In the bulk, these $p$-forms obey the following general action principle

\begin{equation}\label{nuclide-action}
I_\text{Bulk}=\int{d^5x\,\sqrt{-g}\,e^{-\Phi(\zeta)}\,\mathcal{L}_\text{Nuclide}},    
\end{equation}

\noindent where the dilaton field $\Phi(\zeta)$ defines the confinement mechanism, and $\mathcal{L}_\text{Nuclide}$ defines the $p$-form Lagrangian density that sets the dual physics in the bulk. This lagrangian density is defined as 

\begin{multline}\label{p-formaction}
\mathcal{L}_\text{Nuclide}=\frac{\left(-1\right)^p}{2}\,\times\\
\left[\frac{1}{g_p^2}\,g^{mn}\,g^{m_1\,n_1}\ldots\,g^{m_p\,n_p}\nabla_m\,A_{m_1\ldots m_p}\,\nabla_n\,A_{n_1\ldots n
_p}\right. \\
\left.-M_5^2\,g^{m_1\,n_1}\ldots g^{m_p\,n_p}\,A_{m\dots m_p}\,A_{n\ldots n_p}\right].   
\end{multline}

Notice that $g_\text{p}$ is a coupling that sets units in the bulk action. 



We will consider the bulk manifold described by the five-dimensional AdS space parametrized by the Poincaré Patch, defined as 

\begin{equation}
dS^2=\frac{R^2}{\zeta^2}\left(d\zeta^2+\eta_{\mu\nu}\,dx^\mu\,dx^\nu\right),    
\end{equation}

\noindent where $R$ is the AdS curvature Radius and $\zeta$ represents the holographic coordinate and the Greek indices label the minkowskian coordinates. The conformal boundary lies at $\zeta\to0$ as usual. 

After choosing a transverse gauge \cite{Gutsche:2011vb}, i.e.,
\begin{eqnarray}
g^{m_1 m_2}\,A_{m_1\,m_2\ldots m_p}&=&0\\
\nabla^{m_1}\,A_{m_1\,m_2\dots m_p}&=&0,
\end{eqnarray}


\noindent and performing a Fourier decomposition, the action defined above  brings the following set of equations of motion

\begin{equation}\label{master-eq}
\partial_\zeta\left[e^{-B(\zeta)}\,\psi'(\zeta)\right]+M_0^2\,e^{-B(\zeta)}\,\psi(\zeta)-\frac{M_5^2\,R^2}{\zeta^2}\,e^{-B(\zeta)}\,\psi(\zeta)=0,    
\end{equation}

\noindent where we have defined $A_{m_1\ldots m_p}(\zeta,q)=A_{m_1\ldots m_p}(q)\,\psi(\zeta)$, with $A_p(q)$ being the Schwinger source of the operators that create nuclides at the boundary as composite objects and $\psi(\zeta)$ defines the bulk eigenmode. The function $B(\zeta)=\Phi(\zeta)+\beta\,\log\left(R/\zeta\right)$ encloses the geometrical effect associated to the geometry and the confining dilaton, and $\beta=-(3-2\,p)$ fixes the bulk field spin: for scalar fields we have $\beta=-3$ and for vector fields, $\beta=-1$. Notice that we have considered the on-shell mass $-q^2=M_0^2$, defining the nuclide mass. The bulk mass $M_5^2$ defines the \emph{nuclide identity}. 

When we apply the confinement criterium, the normalizable ground state $\psi(\zeta)$ will be dual to the eigenstate associated with the nuclide. The eigenvalue of this bulk $p$-form defines the nuclide mass through the holographic Schrodinger-like potential $V(\zeta)$ constructed by applying the 

\begin{eqnarray}\label{potential-1}
V(\zeta)&=&\frac{1}{4}B'(\zeta)^2-\frac{1}{2}B''(\zeta)+\frac{M_5^2\,R^2}{\zeta^2}.
\end{eqnarray}

This particular picture, provided by the bottom-up models, summarizes the nucleon many-body phenomena in the behavior of geometrically confined bulk fields.  In bottom-up models, the confinement mechanism is defined by the procedure to transform the continuum spectrum into a discrete one. We can do this by deforming the geometry or adding a dilaton profile. 

Following the holography recipe, the conformal dimension of the bulk $p$-form field, i.e., $\Delta$, is dual to the dimension of the operator creating the nuclide, $\text{dim}\,\mathcal{O}$, living at the conformal boundary at $\zeta\to 0$. This is captured into the holographic dictionary as

\begin{equation}
A_p(\zeta,q)\propto A_p(q)\,\zeta^{\Delta-p}.    
\end{equation}

We can write the conformal dimension $\Delta$ in terms of the twist, which accounts for the nuclide constituents, as

\begin{equation}
\Delta\to \text{dim}\,\mathcal{O}=\tau+L,    
\end{equation}

\noindent where $L$ is the angular momentum number.

\textcolor{black}{We will consider nucleons as \emph{constituent objects}. Thus, their associated twist is one. In this context, a given light nucleus should be identified with a \emph{bag} with $N$ nucleons, symmetric under SU$(2)$ as in the Heisenberg Isospin model. Then, the main difference between the bottom-up formulation for QCD and nuclear spectroscopy lies in how we consider the twist. In the former case, twist comes from constituent quarks. In the latter, twist comes from nucleons. As a first approximation, we will consider all nucleons in s-wave, i.e., $L=0$.}

For a general $p$-form, spanned in Fourier space as $A_p(\zeta,q)=A_p(q)\,\psi(\zeta)$, we have in the limit $\zeta \to 0$ that the confinement mechanism is not relevant. Thus, the equations of motion reduce to those in pure AdS: 

\begin{equation}
\left(\frac{\zeta}{R}\right)^{-\beta}\,\partial_\zeta\left[\left(\frac{\zeta}{R}\right)^\beta\,\psi'\left(\zeta\right)\right]+(-q^2) \,\psi(\zeta)-\frac{M_5^2\,R^2}{\zeta^2}\,\psi(\zeta)=0,   
\end{equation}

\noindent where the prime denotes derivative with respect to $\zeta$, the parameter $\beta=-(3-2\,p)$ accounts for $p$-form index effect in the equations, and $M_5^2$ defines the $p$-form bulk mass. 

The solution for this equation is written in terms of Bessel functions of first kind $J_n(x)$ as 

\begin{equation}
\psi(\zeta)=A(q)\,\zeta^{\frac{1-\beta}{2}}\,J_{\left|\Delta-2\right|}\left(M_0\,\zeta\right)    
\end{equation}

and for the bulk mass we have

\begin{eqnarray}\notag
M_5^2\,R^2&=&\left(\Delta-p\right)\left(\Delta-p-1+\beta\right)\\
&=&\left(\Delta-p\right)\left(\Delta+p-4\right).\label{mass5}
\end{eqnarray}

This expression plays a fundamental part since it will modify the behavior of the holographic nuclide ground state in terms of the constituent mass. 

\textcolor{black}{Once we have defined the bulk mass in terms of the conformal dimension, which carries information about nuclide composition and spin, we will write the holographic potential as}

\begin{equation}\label{eq-pot-hol}
    V\left(\zeta\right)=\frac{15-16\Delta+4\Delta^2}{4\,\zeta^2}+\frac{(3-2\,p)}{2\,\zeta}\,\Phi'+\frac{1}{4}\,\Phi'^2-\frac{\Phi''}{2}.
\end{equation}

Solving the potential above, we obtain the nuclide mass spectrum $M_0(Z)$ and the Schrodinger-like modes $\phi_0^Z(\zeta)$ associated with nuclides at the conformal boundary labeled by their atomic number $Z$, recalling that the constituent number is $\Delta=A=2\,Z$ for symmetric nuclides.

\begin{center}
\begin{table*}
\begin{tabular}{||c|c|c||c|c||c|c||c|c||}
\hline
\hline
\multicolumn{9}{||c||}{\textbf{Holographic light nuclide spectrum }}\\
\hline
\hline
\multicolumn{3}{||c||}{\textbf{Experimental nuclide data }} & \multicolumn{2}{c||}{\textbf{Hard Wall Model}} & \multicolumn{2}{c||}{\textbf{Soft Wall Model}}&\multicolumn{2}{c||}{\textbf{Wood-Saxon-like Model}}\\
\hline
\hline
\textbf{$Z$}  &\textbf{Nuclear Spin} &\textbf{$M_0^\text{Exp}$ (u)} & \textbf{$M_0^\text{Th}$ (u)} & \textbf{Rel. Error (\%)}&\textbf{$M_0^\text{Th}$ (u)} & \textbf{Rel. Error (\%)}&\textbf{$M_0^\text{Th}$ (u)} & \textbf{Rel. Error (\%)} \\
\hline
\hline 
2 & 0 & 4.00260 & 4.00260 & 0.00 & 4.002603 & 0.00 & 4.00391 & 0.01 \\
3 & 1 & 6.01511 & 5.91421 & 1.68 & 5.480791 & 8.88 & 6.10883 & 2.34 \\
4 & 0 & 8.00530 & 7.74401 & 3.26 & 8.005206 & 0.01 & 8.16760 & 2.70 \\
5 & 3 & 10.0129 & 9.52800 & 4.84 & 8.372045 & 16.4 & 10.2040 & 2.39 \\
6 & 0 & 12.0000 & 11.2819 & 5.98 & 12.00781 & 3.59 & 12.2212 & 2.21 \\
7 & 1 & 14.0030 & 13.0143 & 7.06 & 13.49952 & 3.59 & 14.2297 & 1.89 \\
8 & 0 & 15.9949 & 14.7303 & 7.91 & 16.01302 & 0.09 & 16.2301 & 1.68 \\
9 & 1 & 18.0009 & 16.4333 & 8.71 & 17.50424 & 2.76 & 18.2244 & 1.40 \\
10& 0 & 19.9949 & 18.1259 & 9.34 & 20.01301 & 0.10 & 20.2137 & 1.23 \\
11& 3 & 21.9944 & 19.8096 & 9.93 & 20.45835 & 6.98 & 22.1989 & 1.02 \\
12& 0 & 23.9850 & 21.4859 & 10.4 & 24.01561 & 0.13 & 24.1805 & 0.89 \\
13& 5 & 25.9869 & 23.1558 & 10.9 & 23.38185 & 10.1 & 26.1593 & 0.72 \\
14& 0 & 27.9769 & 24.8201 & 11.3 & 28.01822 & 0.15 & 28.1374 & 0.61 \\
15& 1 & 29.9783 & 26.4794 & 11.7 & 29.51496 & 1.54 & 30.1092 & 0.47 \\
16& 0 & 31.9721 & 28.1343 & 12.0 & 32.02082 & 0.15 & 32.0811 & 0.36 \\
17& 0 & 33.9738 & 29.7853 & 12.3 & 34.02213 & 0.14 & 34.0512 & 0.24 \\
18& 0 & 35.9675 & 31.4328 & 12.6 & 36.02349 & 0.15 & 36.0196 & 0.15 \\
19& 3 & 37.9691 & 33.0770 & 12.9 & 36.49290 & 3.89 & 37.9867 & 0.05 \\
20& 0 & 39.9626 & 34.7183 & 13.1 & 40.02603 & 0.16 & 39.9523 & 0.03 \\
21& 0 & 41.9655 & 36.3569 & 13.3 & 42.02733 & 0.15 & 41.9168 & 0.12\\
22& 0 & 43.9597 & 37.9929 & 13.6 & 44.02863 & 0.16 & 43.8801 & 0.18\\
23& 0 & 45.9602 & 39.6267 & 13.8 & 46.02994 & 0.15 & 45.8425 & 0.27\\
24& 0 & 47.9542 & 41.9540 & 14.0 & 48.03124 & 0.16 & 47.8038 & 0.33\\
25& 0 & 49.9542 & 42.8880 & 14.1 & 50.03254 & 0.16 & 49.7643 & 0.40\\
26& 0 & 51.9481 & 44.5158 & 14.3 & 52.03384 & 0.16 & 51.7240 & 0.45\\
27& 0 & 53.9484 & 46.1418 & 14.5 & 54.03644 & 0.16 & 53.6829 & 0.51\\
28& 0 & 55.9421 & 47.7663 & 14.6 & 56.03644 & 0.17 & 55.6411 & 0.56\\
29& 1 & 57.9445 & 49.3891 & 14.8 & 57.53525 & 0.71 & 57.5987 & 0.62\\
30& 0 & 59.9418 & 51.0106 & 14.9 & 60.03905 & 0.16 & 59.5555 & 0.67\\
\hline
\hline
\end{tabular}
\caption{This table summarizes the light nuclide spectrum running from $Z=2$ (He) up to $Z=30$ (Ca) symmetric nuclei. We have used $\Lambda_N=0.7794$ u for the hardwall, $\kappa_0=1.0006$ u for the softwall and $A_1=A_2=1.863$ u GeV and $B=2.5$ u for the  Woods-Saxon-like model. Experimental masses read from \cite{Huang:2021nwk}.}
\label{tab:one}
\end{table*}
\end{center}

\subsection{Bottom-up holographic models for the light nuclide masses}
Among the AdS/QCD models, the most successful in describing hadronic properties, particularly hadronic spectra, are the so-called hardwall and softwall models. In such models, the main idea is to place confinement as a geometrical deformation. Confinement in the bulk implies the emergence of bounded states, which will be dual to hadronic states living at the boundary. The dilaton field $\Phi(\zeta)$ (see eqn. \eqref{master-eq} and \eqref{eq-pot-hol} for the holographic potential)  categorizes these bottom-up models.

In the first place, we will consider the so-called hardwall model \cite{Boschi-Filho:2002wdj,Erlich:2005qh}. We fix the dilaton to zero and place a hard cutoff $\Lambda_N$ (a D-Brane) in the holographic coordinate to raise bounded states in a similar form as the infinite square well does in quantum mechanics. In this situation, the spectrum is given in terms of Bessel function zeroes $\alpha_{n,m}$, depending on the hard cutoff, and the number of constituents as:

\begin{equation*}
 M_0(\Delta)=\Lambda_N\,\alpha_{\Delta-2,1}\,\,\,\text{with}\,\Lambda_N=\frac{M_{^4_2\text{He}}}{\alpha_{2,1}}.
\end{equation*}

Masses for the light nuclide spectrum are summarized in table \ref{tab:one}.

Another well-known bottom-up approach we will discuss in this manuscript is the softwall model \cite{Karch:2006pv}. This approach considers setting the dilaton field as  $\Phi(\zeta)=\kappa^2\,\zeta^2$. This choice ensures that the mass spectrum is linear with the excitation number. The dilaton slope $\kappa$ carries information about the quark-antiquark strong interaction. In AdS/QCD, the emergence of the linear spectrum consistent with the Regge theory at the boundary is a clear signal of confinement in bulk. In the case of the softwall model, the quadratic dilaton brings a holographic potential that resembles the 2-dimensional radial harmonic potential with a general linear spectrum given generically by $M_n^2= A\, \kappa^2(n+B)$. Thus, $\kappa$ also defines holographically the Regge slope. 

For the light nuclide case, following the intuition that nuclear force depends on the mass number $A$ and that $\kappa$ carries information about the interaction, an educated guess is to consider that nuclear $\kappa$ scales with the mass number as $\kappa=\sqrt{\frac{\Delta}{2}}\,\kappa_0$. The scale $\kappa_0$ is an energy scale associated with the proton mass that fixes energy units. Thus, the mass spectrum for light nuclides is given by 

\begin{equation}
M_0^2(\Delta)=\Delta\,\kappa_0^2\left(\Delta-p\right). 
\end{equation}

The calculation of the light nuclide masses using the spectrum given above is summarized in table \ref{tab:one}. 

Notice that in both hardwall and softwall, we consider that the constituent number $\Delta$ instead of the excitation number (as in AdS/QCD), fixed to the ground state, defines light nuclide masses. 

In order to quantify the accuracy and precision exhibited by the models, we will follow the RMS analysis. For a model with $N$ parameters used to fit $M$ observables $\mathcal{O}_i$, having relative deviation $\delta\,\mathcal{O}_i$ with the model outcomes, the RMS error is calculated as

\begin{equation}\label{RMS}
\delta_\text{RMS}=\sqrt{\frac{1}{M-N}\,\sum_i^{M}\left(\frac{\delta\,\mathcal{O}_i}{\mathcal{O}_i}\right)^2}.   
\end{equation}

For the hardwall model the RMS error fitting 29 states with one single parameter $\Lambda_N$ is $11.6\,\%$. In the case of the softwall model, the RMS associated with 29 masses fitted with one parameter $\kappa_0$ is $4.4\,\%$.

Hardwall and softwall models accurately describe the light nuclei mass spectrum. However, both potentials have infinite bounded states, holographically dual to light nuclei. In principle, this mass tower is stable and does not decay, implying that heavy nuclei, with large values of $\Delta$, are stable, which phenomenologically is not accurate. \textcolor{black}{Also, the shifting between excited states keeps constant, which is not expected in the nuclear case, where the excited states of a given nuclide are connected with decay processes. These decays do not change the ground mass drastically.} 

Thus, to introduce a finite set of stable ground states regarding the constituent number $\Delta$ \textcolor{black}{with a small enough  mass shifting with the ground state mass}, we will formulate a hybrid holographic potential with a Wood-Saxon-like profile. \textcolor{black}{We will provide a deeper discussion in the next section.} Then, we will reconstruct the associated dilaton associated with this potential. As in the softwall model case, the dilaton will be dependent on the mass number and nuclear spin. \textcolor{black}{It is interesting to notice that Woods-Saxon potential emerges as a good tool to describe nuclear spectra, however, coming from a holographic perspective, different from the original nuclear shell model. Such a model has been quite successful in describing the nuclear structure and provides properties of bound-state and continuum single-particle wavefunctions such as nuclear single-particle energies or nuclear radii calculations. However, WS potential (or any other single-particle) does not help compute total binding energies since it is not based on a specific two-body interaction. This single-particle potential proposed in \cite{PhysRev.95.577} has been used as a less complicated alternative to other multi-particle approaches, as the standard Hartree-Fock calculation (see for example \cite{PhysRev.100.439,Negele:1981tw}). WS-like potential for heavy nuclei provides a good description of nucleon energy levels. Their uses have been extended to other physics branches beyond nuclear physics, such as confined systems in condensed matter \cite{LSCosta_1999}}. 


The key point of the proposed inverse holographic engineering is the dilaton reconstruction.  At sufficient large values of $\zeta$, we expect that the holographic potential acquires a softened profile, flowing asymptotically to a constant value, opposite as in the AdS/QCD soft-wall model, where the potential goes asymptotically to infinite.  The dilaton controls the asymptotic evolution of the holographic potential. Thus, by fixing the asymptotic form of the  potential \eqref{potential-1} with a Woods-Saxon profile, according to the expression  

\begin{equation}\label{dilaton-eqn-diff}
\left[A_1-\frac{A_2}{1+\text{exp}\left(\frac{\zeta-B}{\Delta}\right)}\right]\Delta^2= \frac{(3-2\,p)}{2\,\zeta}\,\Phi'+\frac{1}{4}\,\Phi'^2-\frac{\Phi''}{2}, 
\end{equation}

\noindent \textcolor{black}{we can compute the dilaton field for each nuclide. We have supposed that the depth size in the potential depends on the constituent number encoded in $\Delta$.}  

\textcolor{black}{With this profile we can compute the holographic potential and the holographic light nuclei mass. The table \ref{tab:one} summarizes the numerical results in this model. It is remarkable that the holographic reconstruction of the Woods-Saxon potential provides a very precise model for the nuclear masses. It is a clear case where the dynamics of the bulk modes with respect to the holographic direction captures the spectral properties of the boundary modes dynamics with respect to the radial direction in coordinate space.}

In the Woods-Saxon-like model, having three parameters $A_1$, $A_2$, and $B$ to control the potential well size, modeling 29 light nuclide masses brings an RMS error, following eqn. \eqref{RMS}, around $1.2\,\%$.

\subsection{Holographic nuclide spectra}
\begin{center}
\begin{table*}
\begin{tabular}{||c|c|c||c|c|c||c|c|c||}
\hline
\hline
\multicolumn{9}{||c||}{\textbf{Holographic $_{20}^{40}$Ca excited states}}\\
\hline
\hline 
\multicolumn{3}{||c||}{\textbf{Hardwall}}&\multicolumn{3}{c||}{\textbf{Softwall}}&\multicolumn{3}{c||}{\textbf{Woods-Saxon-like}}\\
\hline
\hline 
\textbf{$n$} & \textbf{$M_n$ (u)} & \textbf{$\Delta\,M_n$(u)}&\textbf{$n$} & \textbf{$M_n$ (u)} & \textbf{$\Delta\,M_n$(u)}&\textbf{$n$} & \textbf{$M_n$ (u)} & \textbf{$\Delta\,M_n$(u)}\\
\hline
\hline 
0 & \textbf{34.7183} & 0 & 0 & \textbf{40.02603} & 0 & 0 & \textbf{39.9523} & 0 \\
1 & 38.8472 & 4.1289 & 1 & 41.0145 & 0.9884 & 1 & 40.0085 & 0.0562 \\
2 & 42.4237 & 7.7054 & 2 & 41.9796 & 1.9536 & 2 & 40.0641 & 0.1117 \\
3 & 45.7299 & 11.012 & 3 & 42.9231 & 2.8971 & 3 & 40.1187 & 0.1665 \\
4 & 48.8703 & 14.152 & 4 & 43.8463 & 3.8203 & 4 & 40.1728 & 0.2206 \\
5 & 51.8971 & 17.179 & 5 & 44.7505 & 4.7244 & 5 & 40.2262 & 0.2740 \\
\hline
\hline 
\end{tabular}
\caption{This table summarizes the $^{40}_{20}$Ca ground state (in bold font) with the first five excited radial states for each holographic model considered. In the hardwall and softwall models, the energy shifting between radial levels and the ground state $\Delta\,M_n=M_n-M_0$ grows with the excitation level beyond the single nucleon mass. In the case of the Woods-Saxon-like approach, the energy shifting is less than the nucleon mass.}
\label{tab:two}
\end{table*}
\end{center}

 
 Let us devote a few comments on the holographic nature of the calculated nuclide spectra, summarized in table \ref{tab:one}. Recall that 
 masses in table \ref{tab:one} are composed by ground states of each holographic potential, characterized by $\Delta$, according to eqn. \eqref{eq-pot-hol}. In the case of hardwall and softwall models coming from AdS/QCD, it is interesting to wonder about the validity of these models applied to the nuclear realm and then extrapolate to the Woods-Saxon-like approach.

In the three models discussed above, the key ingredient is the holographic potential that gives rise to the radial eigenvalue spectrum, i.e., defined in terms of the excitation level. In the hadronic case, the energy shifting between levels is high enough to consider each excitation as a metastable hadronic state. However, the energy shift is not high enough in the nuclear context compared with nuclide mass. Thus, excitation levels in the radial case correspond with the energy transitions between nucleons in the nuclide that, in essence, do not change the nucleus mass far from a nucleon mass, depending on the nature of the energy transition, which would imply changing the atomic number $Z$, leaving the mass number $A$ untouched, or gamma transitions leaving both $Z$ and $A$ intact. These last nuclides, which are not in the ground state, with nucleons in levels above the ground energy, but leaving unchanged the atomic number and mass number, are called \emph{isomers}. Recall that we are considering transitions that leave the nucleon number in the nuclide unaltered. We will consider this last affirmation as a criterion to test the validity of a given holographic model describing nuclear spectroscopy.

Table \ref{tab:two} summarizes the ground state and the first five excited radial states calculated in each holographic model considering the Ca nuclide. In the case of hardwall and softwall models, the energy shifting between radial states and the ground state grows with the excitation number of several nucleon masses. Thus, the excited states cannot be considered the same nuclide described by the ground state. From the nuclear phenomenology, these sorts of transitions are not allowed.


In the case of the Wood-Saxon-like model, the energy shifting between radial level and the ground state is less than 20 $\%$ of the nucleon mass. Thus, in the first five excited states, the nuclide mass does not change beyond one nucleon mass in the case of He and remains almost the same for the Ca and Zn cases. Thus, the Woods-Saxon-like model is more suitable for describing nuclide spectroscopy than the AdS/QCD counterparts at the holographic level. 


%

\section{Nuclear stability}\label{sec4}




Recently, \textcolor{black}{it has been proposed that} 
the configurational entropy (CE) of the hadronic state works as a measure of its stability: the smaller the CE, the more stable the hadron \cite{Bernardini:2016hvx,Bernardini:2016qit,Lee:2017ero,Karapetyan:2021crv}. \textcolor{black}{In the holographic approach, the spectral properties of a given hadron encode in the holographic bulk mode. Thus, bulk modes contain the wave-function holographic necessary to compute the differential configurational entropy of such a hadronic state.} So, it is natural to associate the configurational entropy of the holographic bulk mode describing the nuclear state with its stability. Here we will perform the holographic computation of the CE of the light nuclei in each of the three different holographic models presented in the previous section.

In our case we restrict to bosonic nuclei state, that are integer spin arrangements of many spin $1/2$ baryons. The dual description of bosonic nuclei is encoded in a $p$-form field in the deformed $AdS_5$ space with bulk action given by eq. \eqref{p-formaction}. In this sense, the configuration of the bulk mode appears in the functional dependence on the holographic direction $\zeta$.

Configurational entropy measures the relationship  between the informational content of the physical solutions regarding their equations of motion. CE is also a logarithmic measure of how spatially-localized solutions with given energy content have spatial complexity. Thus, it measures information content in the solutions to the equations of motion. \textcolor{black}{CE has a connection with the relative abundance related to the abundance of the hadronic states. We expect the CE of atomic nuclei to grow with the atomic number.}

In the original formulation, coming from information theory, CE can be interpreted as measuring how much information is necessary to describe localized functions, i.e., e.o.m. solutions,  concerning their parameter set. In general, dynamical solutions come from extremizing an action. CE measures the available information in those solutions.

The association between CE and complexity is translated into stability. Since CE measures the complexity of a given physical system, physical states with higher CE require more energy to be produced in nature than their low CE counterparts. More energy also implies more modes conforming such a physical state, indicating CE increases with the coarseness degree. In this sense, CE is also a measure of stability. Recall that CE measures the relative ordering in field configuration space, showing how energy is related to coarseness. The higher the constituents, the higher the energy and relative configurational entropy. \textcolor{black}{In addition, studies of active matter systems show that the stationary final state of a many-particle system minimizes the configurational entropy \cite{soto2013run,de2021active}. In this sense, dynamical stability is related to configurational entropy: the smaller the CE, the more stable the system.}

Configurational entropy for a discrete variable with probabilities $p_n$ is defined from the Shannon entropy as follows \cite{Gleiser:2012tu,Bernardini:2016qit,Braga:2016wzx}

\begin{equation}
    S_C=-\sum_n\,p_n\,\text{log}\,p_n. 
\end{equation}

In the case of continuous variables, we have  the \emph{differential configurational entropy} (DCE) defined as 

\begin{equation}\label{DCE-def}
S_C\left[f\right]=-\int{d^d\,k\,\tilde{f}\left(k\right)\,\text{log}\,\tilde{f}\left(k\right)},    
\end{equation}

\noindent where $\tilde{f}\left(k\right)=f\left(k\right)/f\left(k\right)_\text{Max}$ defines the modal fraction, $f(k)_\text{Max}$ is the maximum value assumed by $f(k)$. Also we have that $f(k)\in\,L^2\left(\mathbb{R}^2\right)$ i.e., the square-integrable space of functions on the plane. This ensures that $f(k)$ has a defined Fourier transform.  Usually, this $f(k)$ function is associated with the energy density in momentum space, $\rho(k)$. Thus, to compute the DCE for a given physical system, we must address the following algorithm:
\begin{enumerate}
    \item Obtain the localized solutions to the equations of motion.
    \item Evaluate the on-shell energy density.
    \item Transform to momentum space.
    \item Calculate the modal fraction.
    \item Evaluate the DCE integral given in the expression \eqref{DCE-def}.
\end{enumerate}

We will follow this prescription to compute the DCE for the three holographic models discussed above.\\

\begin{center}
\begin{figure*}
  \begin{tabular}{c c c}
    \includegraphics[width=2.4 in]{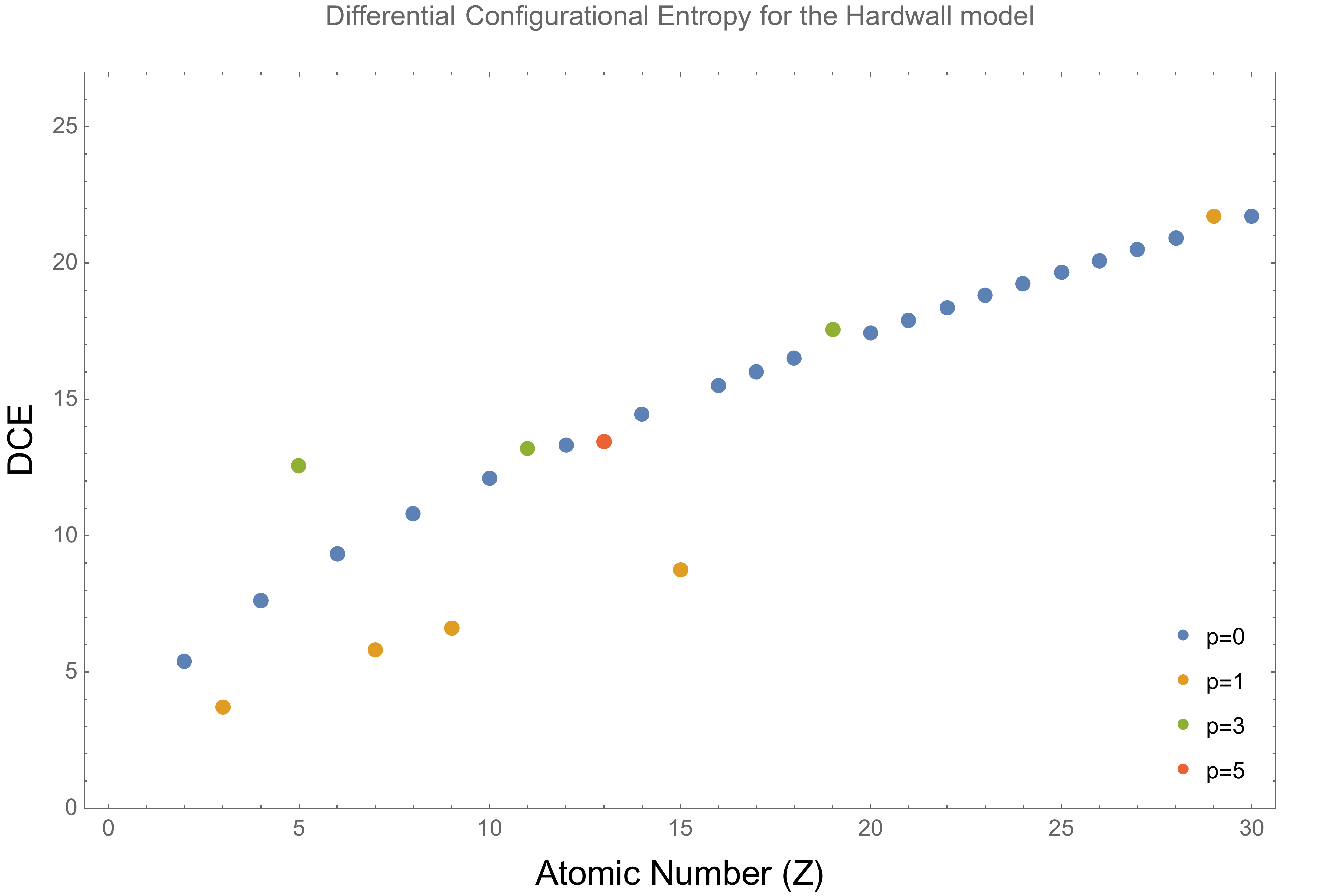}
    \includegraphics[width=2.4 in]{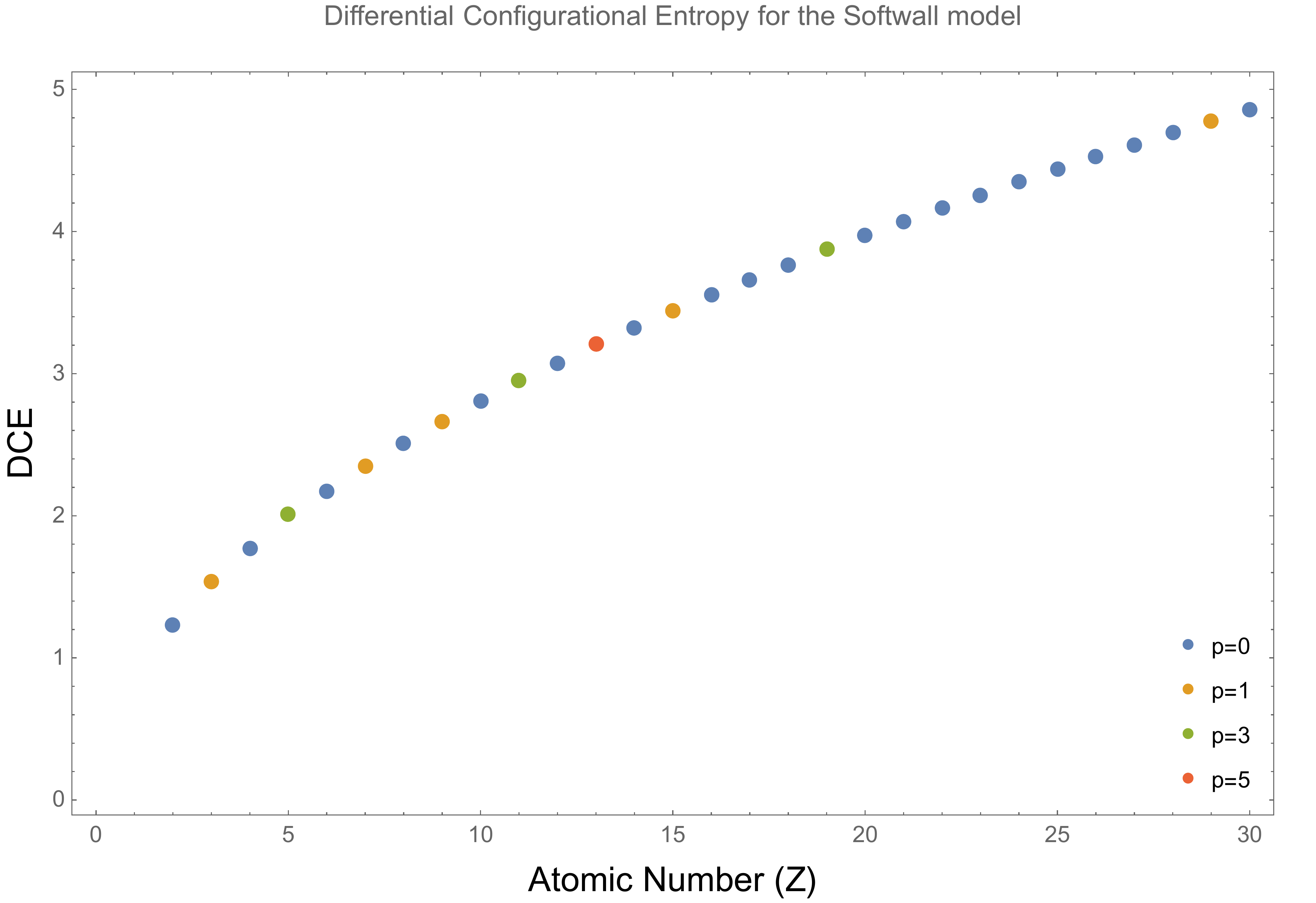}\includegraphics[width=2.4 in]{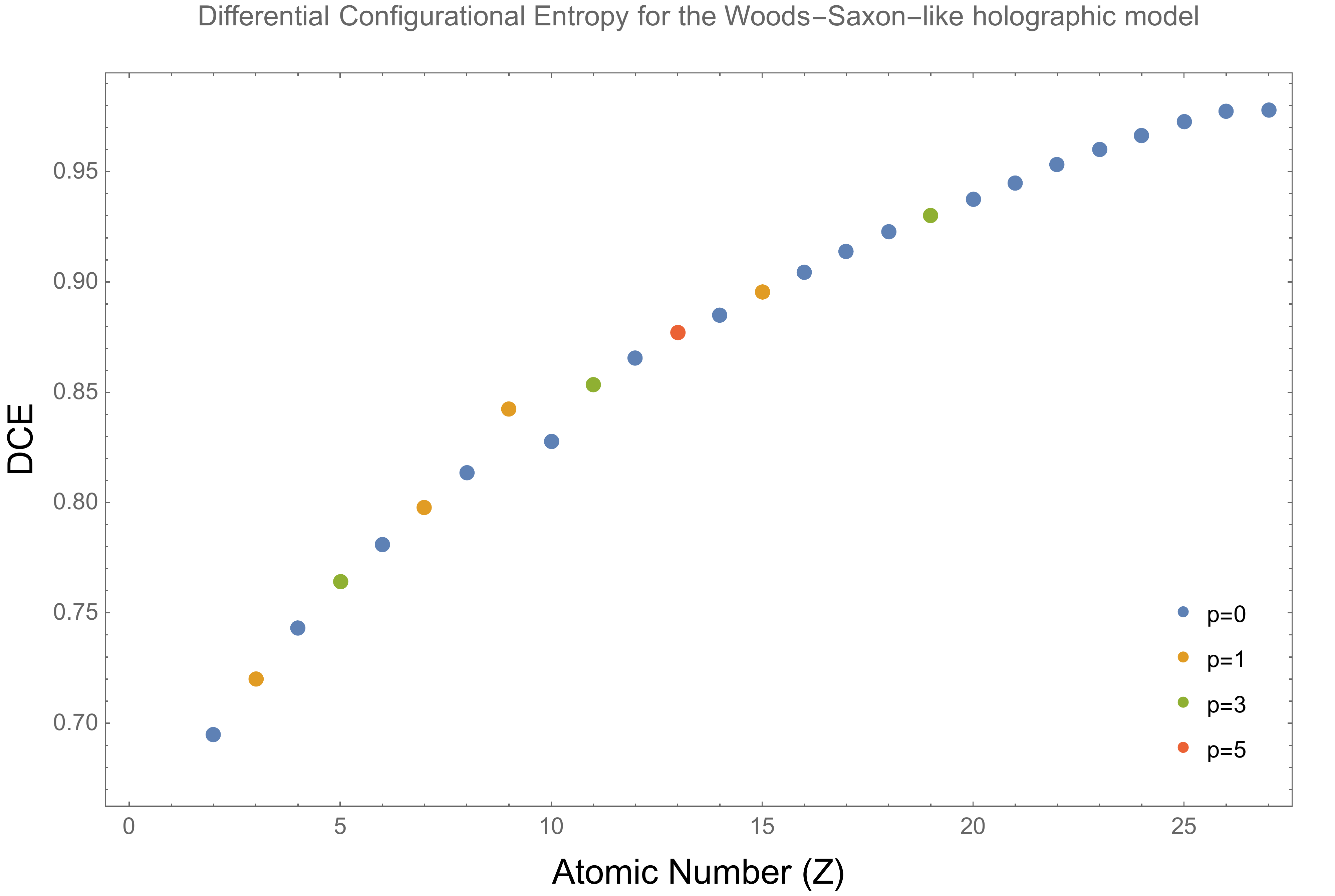}
  \end{tabular}

\caption{Differential Configurational Entropy (DCE) for holographic models considered as a function of the atomic number $Z$. In the left panel we plot the hard wall DCE. The middle panel depicts the DCE for the soft wall model. In the left panel, we depict the Woods-Saxon-like DCE result.}
\label{fig:four}
\end{figure*}
\end{center}

 
 
In the AdS/CFT context, the holographic approach to configurational entropy in bottom-up and top-down AdS/QCD models was made in \cite{Bernardini:2016hvx}. For hadronic states, it was introduced in  \cite{Bernardini:2016qit,Braga:2017fsb,Colangelo:2018mrt,Ferreira:2019nkz,Ferreira:2020iry,daRocha:2021ntm} and references therein. In the context of heavy quarkonium stability, DCE was used as a tool to explore thermal behavior in a colored medium \cite{Braga:2018fyc}, in presence of magnetic fields \cite{Braga:2020hhs} or at finite density \cite{Braga:2020myi}. Recently, in \cite{Braga:2020opg} was used DCE to address the holographic deconfinement phase transition in bottom-up AdS/QCD. 

In our case, we will compute the DCE for the holographic nuclide starting from the associated bulk stress-energy tensor 


\begin{equation}
T_{mn}=\frac{2}{\sqrt{-g}}\,\frac{\partial\left[\sqrt{-g}\,\mathcal{L}_\text{Nuclide}\right]}{\partial\,g^{mn}},    
\end{equation}

\noindent which holds since the action does not depend on metric tensor derivatives. From the action principle \eqref{p-formaction} we can compute the stress-energy tensor for holographic nuclides:











\begin{widetext}
\begin{multline}\label{emtensor}
 T_{mn}=-g_{mn}\,\mathcal{L}_\text{Nuclide}+(-1)^p\,e^{-\Phi} \, \left[\frac{1}{g_p^2}\,g^{m_1\,n_1}\ldots g^{m_p\,n_p}\,\nabla_m\,A_{m_1\ldots m_p}\,\nabla_n\,A_{n_1 \ldots n_p}\right.\\
 \left.+p\,g^{m_2n_2}\ldots g^{m_p n_p}\left(\frac{1}{g^2_p}\,g^{\sigma \rho}\,\nabla_\sigma\,A_{m\,m_2,\ldots m_p}\,\nabla_\rho\,A_{n\,n_2\ldots n_p}-M_5^2 A_{m\,m_2\ldots m_{p}}\,A_{n\,n_2\ldots n_{p}}\right)\right] 
\end{multline}
\end{widetext}

\textcolor{black}{The $p-$form bulk field can be spanned in terms of plane waves as}

\begin{equation}
A_{m_1\ldots m_p}(\zeta,x)=\epsilon_{m_1 \ldots m_p}\,e^{-i q\cdot x}\,\psi(\zeta).    
\end{equation}

\textcolor{black}{Once we define a polarization for the $p-$form field, since nuclides are supposed to be at rest, we will choose a rest frame, i.e., $q=(M_0,\vec{0})$. It is important to remark that eq.(\ref{emtensor}) consider a real p-form field, while the plane wave modes are complex valued. The complex phase is absorbed in the $\Omega$ factor and does not contribut to the DCE. By taking the $00$-component, associated with the energy density $\rho_p(\zeta)$, we obtain}


\begin{multline}\label{pformdensity}
 \rho(\zeta)\equiv T_{00}=\frac{e^{-\Phi(\zeta)}}{2}\left(\frac{\zeta^2}{R^2}\right)^p\times\\
\left\{ \left[\frac{1}{g_p^2}\left(M_0^2\,\psi^2+\psi'^2\right)-\frac{M_5^2\,R^2}{\zeta^2}\psi^2\right] \right\}\,\Omega,
\end{multline}

\noindent where $\Omega$ is a factor carrying plane wave and polarization contraction factors. This factor becomes irrelevant during the modal fraction calculation. 


The Fourier transform reads as

\begin{equation}
    \bar{\rho}(k) = \int_0^{\infty}d\zeta e^{ik\zeta}\rho(\zeta).
\end{equation}

The modal fraction  is defined following \cite{Gleiser:2011di} as

\begin{equation}
   f(k) =\frac{ |\bar{\rho}(k)|^2}{\int dk |\bar{\rho}(k)|^2}.
\end{equation}



The differential configurational entropy for the holographic light nuclide is then written as

\begin{equation}
S_{DCE}=-\int dk \,\tilde{f}(k) \log \,\tilde{f}(k)
\end{equation}

\noindent where $\tilde{f}\left(k\right)=f\left(k\right)/f\left(k\right)_\text{Max}$. The results for the holographic light nuclide DCE, calculated in each model considered, are summarized in the figure \ref{fig:four}. Notice that each nuclide is defined by the ground state calculated from the holographic potential. This localized ground state is characterized by its nuclear spin and mass number encoded into $\Delta$. Thus, although we are not summing over different states, we increase the particle content, i.e., the coarseness degree. Thus, the calculated DCE is a holographic measure of stability.
 
\textcolor{black}{It is essential to make a difference with the hadronic DCE at this stage. The energy density $\rho(\zeta)$, in general,  is a function of the mass spectrum $M^2(n)$ and the particle content. In the hadronic case, the particle content is fixed to the valence quarks while the hadronic mass increases, implying that DCE increases with the excitation number $n$. In the nuclear case considered here, the energy is fixed by the ground state mass while the nucleon content is incremented. The energy density comes from bulk modes calculated from the holographic potential. The potential carries the confinement information inherited by the dilaton field. Thus, in the case of hadrons, the direct consequence is the emergence of Regge trajectories. The bulk mass $ M_5$ controls the particle content in such a hadronic scenario via $\Delta$, which is dual to the dimension of the operators creating hadrons, fixed by the number of valence quarks. In the nuclear case, we extend this idea to consider that the bulk mass carries the nucleon number $A$ (coarseness degree) of the nuclide at hand. In this sense, the DCE measures nuclear stability in the holographic context. However, if the energy and configuration (given by the bulk mass) are degenerate, the CE is also degenerate since there is no quantity in holography that directly measures the inner structure or configuration. According to the holographic dictionary, the state identity (hadronic or nuclear) is defined by the conformal dimension since it is connected with the operator creating these states at the conformal boundary, which does not consider the constituent inner configuration. In this sense, at the holographic level, nuclides can be understood as a bag filled with constituent interacting nucleons.}

In the case of the hardwall model (left panel in figure \ref{fig:four}), DCE does not provide evidence of a single entropy evolution with the atomic number $Z$. Instead, the DCE tends to organize by spin-labeled structures, emulating Regge trajectories for hadrons. This particular behavior comes from the AdS/QCD naturalness that inherited the hardwall since this model was done initially to address hadronic spectra. As was expected, when restricting for only one nuclear spin DCE increases with the nucleon number.

For soft-wall-like model results, summarized in the central panel in figure \ref{fig:four}, all nuclide states define a single trajectory despite their spin, despite their AdS/QCD inherit behavior. DCE increases with the atomic number $Z$, as was expected. We can consider this trajectory a holographic stability line since most of the symmetric light nuclides are stable. Thus, we can expect that at some $Z$, when the nuclide becomes heavier, instability may arise. This observation suggests that asymmetric nuclides can exhibit differences in their relative DCE with their local partners, making them out of the stability trajectory. Also, it suggests that nuclides become unstable for some high values of $Z$. To address the last hypothesis, it is necessary to parametrize the inner nuclide configuration since labeling with particle constituents number only introduces degeneracy, i.e., two different nuclides could have the same mass number.

\textcolor{black}{In the case of the Woods-Saxon-like model, plotted in the right panel of figure \ref{fig:four},  DCE grows with the atomic number, as was expected. We observed a small local jump where spin-1 $_9^{18}$F has a bigger CE than spin-0 $_{10}^{20}$Ne. We do not observe another local jump when running calculations up to $_{28}^{56}$Ni. Neither when we have spin shiftings, as it happens in the neighborhood of $_{19}^{38}$K. Also, as in the softwall model case, the nuclide organization defines a trajectory in terms of stability. However, we can explore this fact further since we do not have a holographic mechanism to describe nucleon configuration inside nuclides. }

\section{Conclusions}\label{conc}
In the present work, we explored three different holographic models for nuclear mass spectroscopy. Motivated by the fact that nuclear force emerges from the strong interaction, we consider the hardwall and softwall models as the first approximation to model light nuclide masses. In both situations, for a fixed nuclide, radial excitations could be interpreted as nuclides different from the ground state since their mass difference is bigger than the proton mass. We propose a different dilaton associated with a holographic Woods-Saxon-like potential to improve this situation. This sort of potential has a bounded above spectrum, consistent with the experimental light nuclide spectrum.

The outcomes we present in table \ref{tab:one} give reasonable results in terms of precision and accuracy. However, if we explore the connection between higher excited states and mass shifts, as table \ref{tab:two} summarizes, we realize that the Woods-Saxon-like holographic reconstruction is the more accurate in reproducing the nuclear mass spectra. 

In our case, the \emph{light nuclide spectrum} is defined as the collection of ground states calculated from holographic potentials, eqn. \eqref{eq-pot-hol}, depending on the constituent number (see table \ref{tab:one}), i.e., the mass number $A$. This situation is different from hadronic physics, where a spectrum is defined as the ground state and their excitations calculated from a single potential with the constituent number fixed. 

We expect that radial excitations do not grow in energy, implying a different nuclide from the ground state (see table \ref{tab:two}). The bulk dilaton should give rise to a holographic potential whose spectrum is bounded from above. In our case, we use a holographic modification of the Woods-Saxon potential. 



\textcolor{black}{Even though the hardwall model does not reproduce the \emph{light nuclide mass spectrum} with high precision, it provides qualitatively correct results. Also, the deviation from the observed spectrum is low compared with experimental data. We have an RMS near 11$\%$. The softwall model also involves only one parameter and has an RMS near 4 $\%$. However, a good description of the nuclear mass spectrum requires a non-trivial dependence of the dilaton coupling on the atomic number, which we implement \emph{ad-hoc}. Indeed, since the dilaton coupling changes with an atomic number, the softwall model considers a one-parameter family of softwall dilatons to describe the nuclear spectra. We also remark that, although hardwall and softwall are models to describe hadrons originally, these models could work, as a first approximation, when describing light nuclei.}


\textcolor{black}{It is interesting to comment about the role played by the dilaton field. Following the softwall original motivation, the dilaton field carries information about the strong interaction nature, translating into the emergence of Regge Trajectories. Thus, our first hypothesis is that the dilaton field can be promoted to the dual object that carries information about confining forces at the boundary. The real question behind this hypothesis relies upon how to build up a proper dilaton to mimic such a confining interaction at the boundary. The answer is in the holographic bottom-up potential depicted in eqn. \eqref{eq-pot-hol}. We followed this path to construct the holographic version of the Woods-Saxon potential, giving the most accurate bottom-up description for nuclide spectroscopy (RMS error near 1 $\%$).} 


\textcolor{black}{In the first case, the excited states make a remarkable difference between hadronic and nuclear spectroscopy since the nature of such states. In the hadronic case, constituents remain the same while the difference between ground and excited states lies in their inner configuration. These configuration patterns define new metastable states different from the ground one, i.e., the mass difference between the excited state and the ground state is not negligible compared to the latter. In the nuclear case, unless we consider SU$(2)$ isospin symmetry,  constituents would change when the ground state moves to excited metastable states. This fact implies that the excited nuclei and the ground state mass difference should not be bigger than one nucleon mass, where transitions are energetically forbidden with more than one nucleon transmuting into another. Thus, a good holographic approach to nuclear spectroscopy should have mass differences between excited and ground states negligible compared to nucleon mass. }

In addition to the spectroscopic analysis, we also perform a stability analysis by considering the configurational entropy. 

\textcolor{black}{In the configurational entropy case, since these sorts of bottom-up models do not include inner nuclear structure, this observable brings clues about stability regarding the number of constituents. Recall that this constituent number information is enclosed in the bulk mass through the conformal dimension of the operator that creates nuclides at the boundary. This operator is written in terms of the twist, carrying constituent information. On the other hand, it is expected that the configurational entropy behavior should characterize the stability in terms of the number of constituents or the atomic number for nuclear systems. Regarding hadrons, configurational  entropy analysis depends on the spin and hadronic mass since it resembles the Regge Trajectory in terms of stability, i.e., higher radial excitations of a given family have bigger configurational entropy than the hadronic ground state. Similar behavior comes when decay constants are introduced in the analysis: higher excitations have lesser decay constant than the ground state. In the nuclear case, configurational entropy analysis shows that the hardwall is not a good holographic model since light nuclides are organized in spin-dependent structures in the DCE plot (see figure \ref{fig:four}). Softwall and Wood-Saxon models have better results since the entropy grows with the constituent number independently from spin.  }

After the spectroscopic and configurational entropic analysis, we conclude that the best among the three models depicted here to describe the light nuclide spectrum is the Woods-Saxon-like bottom-up approach.  

 
\begin{acknowledgments}
We acknowledge the financial support of FONDECYT (Chile) under the grant No. 1180753 (A. V. and M. A. M. C.). Saulo Diles thanks to Fundação Amazônia de Amparo a Estudos e Pesquisas (FAPESPA) for partial financial support.
\end{acknowledgments}
    

\bibliography{apssamp}

\end{document}